\documentclass[twocolumn,showpacs,preprintnumbers,amsmath,amssymb,superscriptaddress]{revtex4}
\usepackage{graphicx}
\usepackage{dcolumn}
\usepackage{bm}

\def\lesssim{\ \raise.3ex\hbox{$<$}\kern-0.8em\lower.7ex\hbox{$\sim$}\ }
\def\gesim{\ \raise.3ex\hbox{$>$}\kern-0.8em\lower.7ex\hbox{$\sim$}\ }
\font\scripti=cmmi7
\font\scriptscripti=cmmi5
\def\sib#1{\setbox0 = \hbox{\scripti #1}
  \kern-.02em\copy0\kern-\wd0
  \kern.04em\box0} 
\def\ssib#1{\setbox0 = \hbox{\scriptscripti #1}
  \kern-.02em\copy0\kern-\wd0
  \kern.04em\box0} 
\font\tenib=cmmib10 
\skewchar\tenib='177 \skewchar\tenib='177 \skewchar\tenib='177
\def\pbold#1{\setbox0 = \hbox{$ #1 $}
  \kern-.022em\copy0\kern-\wd0
  \kern.011em\copy0\kern-\wd0
  \kern.011em\copy0\kern-\wd0
  \kern.011em\copy0\kern-\wd0
  \kern.011em\box0} 

\usepackage{graphicx}
\usepackage{dcolumn}
\usepackage{bm}

\begin{document}

\title{A possible method to confirm $\pm s$-wave pairing state using the Riedel anomaly in Fe-pnictide superconductors}

\author{Daisuke Inotani$^{1}$ and Yoji Ohashi$^{1,2}$}
 \affiliation{$^{1}$Department of Physics, Keio University, 3-14-1 Hiyoshi, Kohoku-ku, Yokohama 223-8522, Japan \\
 $^{2}$CREST(JST), 4-1-8 Honcho, Saitama 332-0012, Japan}

\date{\today}

\begin{abstract}
We theoretically propose a method to identify $\pm s$-wave order parameter in recently discovered Fe-pnictide superconductors. Our idea uses the Riedel anomaly in ac-Josephson current through an SI($\pm$ S) (single-band $s$-wave superconductor/insulator/$\pm$s-wave two-band superconductor) junction. We show that the Riedel peak effect leads to vanishing ac-Josephson current at some values of biased voltage.  This phenomenon does not occur in the case when the $\pm s$-wave superconductor is replaced by a conventional $s$-wave one, so that the observation of this vanishing Josephson current would be a clear signature of $\pm s$-wave pairing state in Fe-pnictide superconductors. 

\end{abstract}

\maketitle

In the current stage of research on Fe-pnictide superconductors\cite{Kamihara,Chen1,Chen2,Ren,Zhian}, the symmetry of superconducting order parameter is one of the most important issues. Since the discovery of superconductivity in LaFeAsO$_{1-x}$F$_x$\cite{Kamihara}, great experimental and theoretical efforts have clarified various key properties of these materials. FeAs-layers form a quasi-two dimensional electron system, consisting of hole and electron pockets around the $\Gamma$- and $M$-point, respectively\cite{Leb,Singh,Boeri,Kuroki,Cao,Ding,Kondo,Liu}. An antiferromagnetic (AF) phase exists without carrier doping\cite{Cruz}, so that the possibility of pairing mechanism associated with AF spin fluctuations has been discussed\cite{Mazin,Wang,Bang,Korshunov}. The decrease of Knight shift\cite{Grafe} below the superconducting phase transition temperature $T_{\rm c}$ indicates a singlet pairing state. A tunneling experiment\cite{Wang2}, as well as angle-resolved photoemission spectroscopy (ARPES)\cite{Ding,Kondo}, have shown that Fe-pnictides are multigap superconductors. The ARPES experiment also reports that the order parameter in each band may have a nodeless $s$-wave symmetry\cite{Ding,Kondo}. While this is consistent with the exponential temperature dependence of the penetration depth far below $T_{\rm c}$\cite{Hashimoto}, it seems contradicting with the $T^3$-behavior of NMR-$T_1^{-1}$\cite{Ishida1,Ishida2}, implying the existence of nodes.  
\par
As a candidate for the symmetry of order parameter in Fe-pnictide superconductors, a $\pm s$-wave state has been recently proposed\cite{Mazin,Wang,Bang,Korshunov}. In this pairing state, nodeless $s$-wave order parameters in electron and hole bands have opposite sign to each other. This unconventional superconductivity has been shown to consistently explain the observed superconducting properties mention above\cite{Grafe,Wang2,Ding,Kondo}, except for the power-law behavior of NMR-$T_1^{-1}$\cite{Ishida1,Ishida2}. However, some theory groups have pointed out that the NMR result can be also explained within the framework of $\pm s$-wave superconductivity, when one includes impurity scattering\cite{Parker} and/or anisotropic Fermi surfaces\cite{Machida}. It has been also reported that the enhancement of inelastic neutron scattering rate at a finite momentum transfer observed in superconducting Ba$_{0.6}$K$_{0.4}$Fe$_2$As$_2$ is consistent with the $\pm s$-wave scenario\cite{Korshunov,Chris}. Since a model calculation including a pairing interaction mediated by AF spin fluctuations supports $\pm s$-wave superconductivity\cite{Bang}, confirming the $\pm s$-wave order parameter in Fe-pnictides is also crucial for clarifying the pairing mechanism of these materials.
\par
\begin{figure}
\centerline{\includegraphics[width=4.5cm,height=2.5cm]{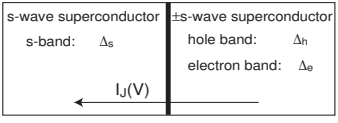}}
\caption{Model SI($\pm$S)-junction which we consider in this paper. The left of the junction is a single-band (denoted by s-band) $s$-wave superconductor with the order parameter $\Delta_s$. The $\pm s$-wave superconductor on the right of the junction has two bands denoted by h-band and e-band, with the order parameter $\Delta_h$ and $\Delta_e$, respectively. $I_J$ is the Josephson current through the junction.
}
\label{fig1}
\end{figure}

In this paper, we theoretically propose a method to confirm the $\pm s$-wave order parameter in Fe-pnictide superconductors. In identifying the pairing symmetry of unconventional superconductivity, phase-sensitive experiments are very powerful. For example, the so-called $\pi$-junction SQUID played crucial roles to identify the $d_{x^2-y^2}$-wave order parameter in high-$T_{\rm c}$ cuprates\cite{Harlingen}. Our idea uses the ac-Josephson current $I_J$ through an SI($\pm$S) (single-band $s$-wave superconductor/insulator/$\pm s$-wave superconductor) junction shown in Fig.\ref{fig1}. In this case, $I_J$ consists of two components associated with two bands in the $\pm s$-wave superconductor. Because of the sign difference of two order parameters in the $\pm s$-wave state, these two current components are found to flow in the opposite direction to each other.  In addition, as in the case of ordinary ac-Josephson current, each current component shows the Riedel anomaly\cite{Riedel,Hamilton}, where the Josephson current diverges at a certain value of biased voltage $V$ across the junction. These two phenomena are shown to give vanishing total ac-Josephson current $I_J$ at some values of $V$.  This vanishing $I_J$ does {\it not} occur when the order parameters in the two-band superconductor have the same sign. Since the ARPES experiment reports a nodeless $s$-wave order parameter in each band\cite{Ding,Kondo}, the observation of the vanishing ac-Josephson current would be a clear signature of $\pm s$-wave state in Fe-pnictides. 

To explain details of our idea, we explicitly calculate the ac-Josephson current through the SI($\pm$S)-junction in Fig.\ref{fig1}. The Hamiltonian is given by
\begin{equation}
H=H_{s}+H_{\pm s}+H_T,
\label{eq.1}
\end{equation}
where $H_s$ and $H_{\pm s}$, respectively, describe the single-band $s$-wave superconductor on the left of the junction and $\pm s$-wave superconductor on the right of the junction. Tunneling effects are described by $H_T$. In the BCS approximation, $H_s$ is given by
\begin{equation}
H_s=\sum_{{\bf p},\sigma}\varepsilon_{\bf p}^s 
a_{{\bf p}\sigma}^{s\dagger} a_{{\bf p}\sigma}^s
+\sum_{\bf p}
\Bigl[
\Delta_s a_{{\bf p}\uparrow}^{s\dagger} a_{-{\bf p}\downarrow}^{s\dagger}+h.c.
\Bigr].
\label{eq.2}
\end{equation}
Here, $a_{{\bf p}\sigma}^{s\dagger}$ is the creation operator of an electron in the s-band with the kinetic energy $\varepsilon_{\bf p}^s$, measured from the Fermi energy. $\displaystyle \Delta_s=U_s\sum_{\bf p}\langle a_{-{\bf p}\downarrow}^sa_{{\bf p}\uparrow}^s\rangle$ is the order parameter in the $s$-band, where $U_s<0$ is a pairing interaction. 
\par
For $H_{\pm s}$ in Eq. (\ref{eq.1}), we simply assume a two-band system as a minimal model to describe $\pm s$-wave superconductivity (although band calculations\cite{Leb,Singh,Boeri,Kuroki,Cao,Mazin}, as well as ARPES experiment\cite{Ding}, indicate the existence of more than two bands).  We also do not discuss the origin of the pairing interaction in this paper, but simply employ the following model Hamiltonian\cite{Bang,Shul}
\begin{eqnarray}
&H_{\pm s}&
=\sum_{{\bf p},\sigma,\alpha=e,h} \varepsilon_{\bf p}^{\alpha} c_{{\bf p}\sigma}^{\alpha \dagger} c_{{\bf p} \sigma}^{\alpha}
\nonumber
\\
&+&
\sum_{{\bf p},{\bf p}',{\bf q} \atop \alpha,\alpha'=e,h}U_{\alpha\alpha'} 
c_{{\bf p}+{\bf q}/2 \uparrow}^{\alpha \dagger}
 c_{-{\bf p}+{\bf q}/2 \downarrow}^{\alpha \dagger} 
 c_{-{\bf p}'+{\bf q}/2 \downarrow}^{\alpha'} 
 c_{{\bf p}'+{\bf q}/2 \uparrow}^{\alpha'},
\nonumber
\\
\label{eq.3}
\end{eqnarray}
where $c_{{\bf p}\sigma}^{\alpha\dagger}$ is the creation operator of an electron in the $\alpha (={\rm e,h})$-band, with the kinetic energy $\varepsilon_{\bf p}^\alpha$, measured from the Fermi level. $U_{\alpha\alpha}$ is an intraband interaction in the $\alpha$-band. $U_{\alpha\alpha'}$ ($\alpha\ne\alpha'$) describes a pair tunneling between the e-band and h-band. (We take $U_{\rm eh}=U_{\rm he}$.) In the mean-field approximation, Eq. (\ref{eq.3}) reduces to
\begin{eqnarray}
H_{\pm s}=\sum_{{\bf p},\sigma,\alpha} \varepsilon_{\bf p}^{\alpha} c_{{\bf p} \sigma}^{\alpha \dagger} c_{{\bf p} \sigma}^{\alpha} 
+\sum_{{\bf p},\alpha}
\Bigl[\Delta_\alpha c_{{\bf p} \uparrow}^{\alpha \dagger}
 c_{-{\bf p} \downarrow}^{\alpha \dagger}+h.c.
 \Bigr].
\label{eq.4}
\end{eqnarray}
Here, the order parameters $\Delta_h$ and $\Delta_e$ are given by
\begin{eqnarray}
\Delta_{h} = U_{hh}\sum_{\bf p} \langle c_{-{\bf p} \downarrow}^{h} c_{{\bf p} \uparrow}^{h} \rangle 
+ U_{he}\sum_{\bf p} \langle c_{-{\bf p} \downarrow}^{e} c_{{\bf p} \uparrow}^{e} \rangle,
\label{eq.5}
\end{eqnarray}
\begin{equation}
\Delta_{e} = U_{ee}\sum_{\bf p}  \langle c_{-{\bf p} \downarrow}^{e} c_{{\bf p} \uparrow}^{e} \rangle 
+ U_{he}\sum_{\bf p} \langle c_{-{\bf p} \downarrow}^{h} c_{{\bf p} \uparrow}^{h} \rangle.
\label{eq.6}
\end{equation} 
\par
The $\pm s$-wave state is easily obtained, when one sets $U_{he}>0$ and $U_{ee}=U_{hh}=0$. In this case, Eqs. (\ref{eq.5}) and (\ref{eq.6}) give the coupled equations,
\begin{equation}
\Delta_h=-U_{he}
\sum_{\bf p}{\Delta_e \over 2\sqrt{{\varepsilon_{\bf p}^e}^2+|\Delta_e|^2}}
\tanh{1 \over 2T}\sqrt{{\varepsilon_{\bf p}^e}^2+|\Delta_e|^2},
\label{eq.7}
\end{equation}
\begin{equation}
\Delta_e=-U_{he}
\sum_{\bf p}{\Delta_h \over 2\sqrt{{\varepsilon_{\bf p}^h}^2+|\Delta_h|^2}}
\tanh{1 \over 2T}\sqrt{{\varepsilon_{\bf p}^h}^2+|\Delta_h|^2}.
\label{eq.8}
\end{equation}
These equations have solutions only when the sign of $\Delta_h$ is opposite to the sign of $\Delta_e$. We note that $\Delta_h$ and $\Delta_e$ have the same sign when $U_{he}<0$. We also note that the h-band and e-band have the same $T_{\rm c}$, given by\cite{Shul}
\begin{equation}
T_{\rm c}={2\gamma \omega_c \over \pi}
e^{-{1 \over |U_{he}|\sqrt{N_e(0)N_h(0)}}},
\label{eq.9}
\end{equation}
where $\gamma=1.78$, and $\omega_c$ is the ordinary cutoff energy in the BCS theory. $N_\alpha(0)$ is the density of states at the Fermi level in the normal state of the $\alpha$-band. 
\par
\begin{figure}
\centerline{\includegraphics[width=5cm,height=3.5cm]{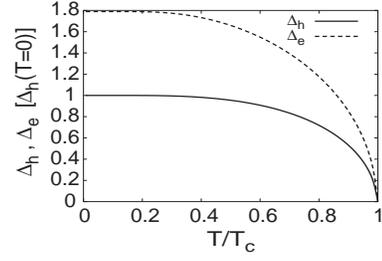}}
\caption{Temperature dependence of the order parameter $|\Delta_{h}|$ and $|\Delta_{e}|$ obtained from the coupled equations (\ref{eq.6}) and (\ref{eq.7}), normalized by the value $\Delta_h(T=0)$. We set $U_{hh}=U_{ee}=0$, $|U_{he}|(N_h(0)+N_e(0))=1.0$, and $N_{h}(0)/N_{e}(0)=0.4$, where $N_\alpha(0)$ ($\alpha$=e,h) is the density of states at the Fermi level in the normal state of the $\alpha$-band. $\Delta_h$ and $\Delta_e$ have opposite sign to each other when $U_{he}>0$, while they have the same sign when $U_{he}<0$.
\label{fig2}}
\end{figure}

Figure \ref{fig2} shows the calculated $\Delta_h$ and $\Delta_e$ from Eqs. (\ref{eq.7}) and (\ref{eq.8}). We will use these results in evaluating the ac-Josephson current. In this regard, we briefly note that, although we take $U_{ee}=U_{hh}=0$ to realize $\pm s$-wave superconductivity in a simply manner, the following discussions on the ac-Josephson effect is not affected by detailed values of $U_{\alpha\alpha'}$, as far as $\pm s$-wave state is realized. 

\par 
The tunneling Hamiltonian in Eq. (\ref{eq.1}) has the form $H_T=A+A^\dagger$, where
\begin{equation}
A = \sum_{{\bf p},{\bf k}\sigma \atop \alpha=h,e} T_{{\bf p},{\bf k}}^{\alpha} a_{{\bf p}\sigma}^{s\dagger} c_{{\bf k}\sigma}^{\alpha}\equiv\sum_\alpha A_\alpha. 
\label{eq.10}
\end{equation}
Here, $T_{{\bf p},{\bf k}}^\alpha$ is the tunneling matrix element between the s-band and $\alpha$-band, which satisfies the time reversal symmetry, as $T_{{\bf p},{\bf k}}^{\alpha} = T_{-{\bf p},-{\bf k}}^{\alpha*}$. Assuming a weak junction, we calculate the tunneling current $I\equiv -e\langle \dot{N}_s\rangle=ie(A-A^\dagger)$ within the lowest order in terms of $T_{{\bf p},{\bf k}}^\alpha$ (where $N_s=\sum_{{\bf p},\sigma}a_{{\bf p}\sigma}^{s\dagger}a^s_{{\bf p}\sigma}$ is the total number operator of electrons on the left of the junction in Fig.\ref{fig1}). Then, we find
\begin{eqnarray}
I(t) = e \int_{-\infty}^t dt' \langle [A(t)-A^\dagger(t),A(t')+A^\dagger(t')] \rangle_0.
\label{eq.11}
\end{eqnarray}
Here, the statistical average $\langle\cdot\cdot\cdot\rangle_0$ is taken in the absence of $H_T$, and $A(t)\equiv e^{i(H_s+H_{\pm s})t}A e^{-i(H_s+H_{\pm s})t}$. 
\par
Effects of finite voltage $V$ across the junction is conveniently incorporated into Eq. (\ref{eq.11}) by replacing $A(t)$ by $e^{-ieVt}A(t)$. Equation (\ref{eq.11}) involves both the Josephson current $I_J$ and quasi-particle current $I_q$. Extracting the former component, we find that $I_J$ consists of the tunneling current between the s- and h-band, and that between s- and e-band, as 
\begin{equation}
I_J=-2e\sum_{\alpha=h,e}{\rm Im}
\Bigl[e^{-2ieVt}\Pi_\alpha(\omega=eV)\Bigr],
\label{eq.12}
\end{equation}
where 
\begin{equation}
\Pi_\alpha(\omega)=-i\int_{-\infty}^t dt e^{i\omega t}
\langle [A_\alpha(t),A_\alpha(0)]\rangle_0.
\label{eq.13}
\end{equation}
Equation (\ref{eq.13}) can be calculated from the analytic continuation of the corresponding thermal Green's function
\begin{eqnarray}
&\Pi&_\alpha(i\nu_n)
=
-\int_0^{1/T} d\tau e^{i\nu_n\tau}
\langle T_\tau\{A_\alpha(\tau)A_\alpha(0)\}\rangle_0
\nonumber
\\
&=&
-2T\sum_{{\bf p},{\bf k}}|T^\alpha_{{\bf p},{\bf k}}|^2 
\sum_{\omega_m} G^s_{21}({\bf p},i\omega_m)G_{12}^\alpha({\bf k},i\omega_m+i\nu_n),
\nonumber
\\
\label{eq.14}
\end{eqnarray}
where $\nu_n$ and $\omega_m$ are the boson and fermion Matsubara frequencies, respectively, and $A(\tau)\equiv e^{\tau(H_s+H_{\pm s})}Ae^{-\tau(H_s+H_{\pm s})}$. The off-diagonal Green's function is given by
\begin{equation}
G^\lambda_{12}({\bf p},i\omega_m)=-{\Delta_\lambda \over \omega_m^2+\varepsilon_{\bf p}^{\lambda2}+|\Delta_\lambda|^2}~~~(\lambda=s,\alpha),
\label{eq.15}
\end{equation}
which satisfies $G^\lambda_{21}=G_{12}^{\lambda*}$.

\begin{figure}
\centerline{\includegraphics[width=5.5cm,height=7cm]{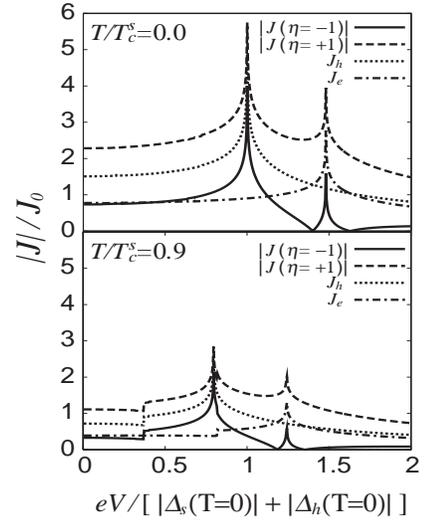}}
\caption{Calculated ac-Josephson current $|J|$, normalized by $J_{0} \equiv \sqrt{G_{h}G_{e}}[|\Delta_{h}(T=0)| + |\Delta_{e}(T=0)|]/e$, as a function of biased voltage $V$. (a) $T=0$. (b) $T/T_{\rm c}^s=0.9$, where $T_{\rm c}^s$ is $T_{\rm c}$ of the superconductor on the left of the junction in Fig.\ref{fig1}. The Riedel anomaly can be seen at $eV/(|\Delta_s|+|\Delta_h|) = 1.0$ and 1.48 in panel (a), and at 0.80 and 1.24 in panel (b). In addition to these peaks, we also find weak singularities at $eV=|\Delta_\alpha|-|\Delta_s|$, for example, $eV/(|\Delta_s|+|\Delta_h|)\simeq 0.37$ in panel (b). We take $\langle T_{{\bf p},{\bf k}}^h\rangle=\langle T_{{\bf p},{\bf k}}^e\rangle$, for simplicity. For the values of $\Delta_h(T)$ and $\Delta_e(T)$, the results in Fig.\ref{fig2} are used. Values of the interaction $U_s$ is chosen as to realize $\Delta_h(T=0)/\Delta_s(T=0)=1.5$. 
\label{fig3}}
\end{figure}

\par
For simplicity, we approximate the tunneling matrix element $T_{{\bf p},{\bf k}}^\alpha$ to the value averaged over the Fermi surface ($\equiv \langle T_{{\bf p},{\bf k}}^\alpha\rangle$). Executing the momentum summations in Eq. (\ref{eq.14}), we obtain $\Pi_\alpha(i\nu_n)=2\pi^2|\langle T_{{\bf p},{\bf k}}^\alpha\rangle|^2N_s(0)N_\alpha(0)\Delta_s^*\Delta_\alpha\Lambda(i\nu_n)$, where
\begin{equation}
\Lambda(i\nu_n)=T\sum_{\omega_m}
{1 \over \sqrt{\omega_m^2+|\Delta_s|^2}}
{1 \over \sqrt{(\omega_m+\nu_n)^2+|\Delta_\alpha|^2}}.
\label{eq.17}
\end{equation}
Here, $N_s(0)$ is the density of states at the Fermi level in the normal state of the s-band. As usual, we evaluate the $\omega_m$-summation in Eq. (\ref{eq.17}) by transforming it into the complex integration. Changing the integration path so as to be able to carry out the analytic continuation in terms of $i\nu_n$, we execute $i\nu_z\to\omega+i\delta$. Substituting the result into Eq. (\ref{eq.12}), we find that the sine-component of ac-Josephson current ($\equiv {\bar I}_J$) can be written as ${\bar I}_J=J\sin(2eV+\phi_s-\phi_h)$, where $\phi_s$ and $\phi_h$ are the phases of the order parameter $\Delta_s$ and $\Delta_h$, respectively. The coefficient $J$ has the form
\begin{equation}
J=J_{h}+\eta J_{e},
\label{eq.18}
\end{equation}
where $J_h$ and $J_e$ come from the tunneling current between the s-band and e-band and that between s-band and h-band, respectively. They are given by\cite{Larkin}
\begin{eqnarray}
J_\alpha
&=&
{G_\alpha \over e}{|\Delta_s||\Delta_\alpha| \over 2}
\int_{-\infty}^\infty dz \tanh{|z| \over 2T}
\nonumber\\
&\times&
\biggl[
{
\theta(|\Delta_s|-|z-eV|)\theta(|z|-|\Delta_\alpha|) 
\over 
\sqrt{|\Delta_s|^2-(z-eV)^2}\sqrt{z^2-|\Delta_{\alpha}|^2}
}
\nonumber
\\
&+&
{
\theta(|z|-|\Delta_s|)\theta(|\Delta_\alpha|-|z+eV|) 
\over 
\sqrt{z^2-|\Delta_s|^2}\sqrt{|\Delta_{\alpha}|^2-(z+eV)^2}
}
\Biggr],
\label{eq.19}
\end{eqnarray}
where $G_\alpha=4\pi e^2N_s(0)N_\alpha(0)|\langle T_{{\bf p},{\bf k}}^\alpha\rangle|^2$. In Eq. (\ref{eq.18}), $\eta$ involves useful information about the phase difference between $\Delta_h=|\Delta|^{i\phi_h}$ and $\Delta_e=|\Delta_e|e^{i\phi_e}$, as
\begin{eqnarray}
\eta=
\left\{
\begin{array}{ll}
-1& ~~~~~(\phi_e=\phi_h+\pi), \\
+1& ~~~~~(\phi_e=\phi_h).
\end{array}
\right.
\label{eq.20}
\end{eqnarray}
\par
When $\eta=-1$ (SI($\pm$ S)-junction), the phase difference between $\Delta_h$ and $\Delta_e$ equals $\pi$. In this case, the current $J_e$ flows in the opposite direction to $J_h$. This leads to the suppression of the total Josephson current as $J=J_h-J_e$. In contrast, when $\phi_e=\phi_h$ in the case of $\eta=+1$, the Josephson current is simply given by the sum of two current components, as $J=J_h+J_e$.
\par
Figure \ref{fig3}(a) shows the magnitude of ac-Josephson current $J$ at $T=0$, as a function of biased voltage $V$. Each component $J_h$ and $J_e$ has a peak at $eV=|\Delta_s|+|\Delta_\alpha|$ ($\alpha=e,h$) (Riedel anomaly). The resulting total ac-Josephson current $J(\eta=-1)=J_h-J_e$ vanishes when the voltage $V$ satisfies $J_h(eV)=J_e(eV)$. (See the solid line in Fig.\ref{fig3}(a).) The Riedel peaks at $eV=|\Delta_s|+|\Delta_h|$ and $eV=|\Delta_s|+|\Delta_e|$ guarantee that this condition is always satisfies at a voltage ($\equiv V_0$) in the region\cite{note},
\begin{equation}
|\Delta_s|+{\rm Min}[|\Delta_h|,|\Delta_e|]<eV_0<|\Delta_s|+{\rm Max}[|\Delta_h|,|\Delta_e|].
\label{eq.21}
\end{equation}
In contrast, when the phase difference between $\Delta_h$ and $\Delta_e$ is absent ($\eta=+1$), $J$ does not vanish, but is always finite. (See the dashed line in Fig.\ref{fig3}(a).) Thus, the observation of the vanishing ac-Josephson current would be a clear signature of $\pm s$-wave state in Fe-pnictides.
\par
The vanishing Josephson current can be also seen at finite temperatures, as shown in Fig.\ref{fig3}(b). On the other hand, when $|\Delta_h|=|\Delta_e|$ and $J_h(eV)\ne J_e(eV)$ are accidentally satisfied, the Riedel peaks in $J_h$ and $J_e$ appear at the same value of $V$, so that $J(\eta=-1)=J_h-J_e$ only has one Riedel peak at $eV=|\Delta_h|+|\Delta_s|$. In this special case, the vanishing $J$ is not obtained even in the SI($\pm$S)-junction. However, since two different energy gaps have been observed in Fe-pnictide superconductors\cite{Wang2,Ding,Kondo}, we can determine the relative sign of the two order parameters corresponding to the observed two energy gaps by our method.
\par
One can immediately extend our idea to the case with more than two order parameters. In this case, when all the order parameters do not have the same sign, we again obtain the vanishing ac-Josephson current due to the same mechanism discussed in this paper.
\par
Multiband superconductivity is affected by even non-magnetic impurities\cite{Ohashi}, so that the Riedel anomaly may be weakened by impurity effects. The suppression of the Riedel anomaly is also expected when one includes anisotropic Fermi surfaces. When the Riedel peak in $J_e$ is broadened and the peak height becomes smaller than the value of $J_h$ at $eV=|\Delta_s|+|\Delta_e|$ in Fig.\ref{fig3}(a), the vanishing ac-Josephson current is no longer obtained. In this case, however, unless the Riedel peak becomes very broad, $J$ would show a dip (peak) structure at $eV=|\Delta_s|+|\Delta_e|$ when $\eta=-1$ ($\eta=+1$), which may be still useful to confirm $\pm s$-wave superconductivity. Since any real superconductor more or less has impurities, as well as anisotropic band structure, it is an interesting problem how our idea discussed in this paper is modified when more realistic situations are taken into account. We will separately discuss this problem in our future paper.
\par
To conclude, we have studied a possible method to confirm the $\pm s$-wave pairing symmetry in recently discovered Fe-pnictide superconductors. Using the Riedel anomaly and the fact that the ac-Josephson current through the SI($\pm$S)-junction consists of two components flowing toward the opposite direction to each other, we obtained the vanishing ac-Josephson current at certain values of biased voltage. This phenomenon is absent when the $\pm$s-wave superconductor is replaced by an $s$-wave superconductor where the order parameters have the same sign. Since the symmetry of order parameter is deeply related to the mechanism of superconductivity, our method discussed in this paper would be also helpful in clarifying the mechanism of superconductivity in Fe-pnictides.
\par
We would like to thank S. Tsuchiya and R. Watanabe for useful discussions. This work was supported by a Grant in Aid from Mext, and the CTC program of Japan.



\begin{thebibliography}{99}
\bibitem{note} In Fig.\ref{fig3}, in addition to $V_0$ which satisfies Eq. (\ref{eq.21}), one also obtains the vanishing Josephson current at $eV\gesim|\Delta_s|+{\rm Max}[|\Delta_h|,|\Delta_e|]$.
\end{thebibliography}
\end{document}